\documentstyle[twocolumn,prc,aps,psfig]{revtex}

\begin{document}

\draft \title{The $\pi NN$ vertex function in a meson-theoretical model
}

\author {R. B\"ockmann$^{1,3}$, C. Hanhart$^{2}$, O. Krehl$^{1}$,
S. Krewald$^{1}$, and J. Speth$^{1}$}

\address {1)Institut f\"{u}r Kernphysik, Forschungszentrum J\"{u}lich GmbH, \\
D-52425 J\"{u}lich, Germany \\
2)Department of Physics and INT, University of Washington, Seattle, \\
WA 98195, USA \\
3)present address: MPI f\"ur biophysikalische Chemie, Am Fa\ss{}berg 11\\ 37077 G\"ottingen \\ }

\date{\today}

\maketitle

\begin{abstract}
The $\pi NN$ vertex function is calculated within a
dispersion theoretical approach, including both
$\pi\rho$ and $\pi\sigma$ intermediate states,
where the $\sigma$ meson is an abbreviation for a
correlated pion pair in a relative S-wave with
isospin I=0.
A strong coupling between
the $\pi\sigma$ and $\pi\rho$ states is found. This leads to
a softening of the   $\pi NN$ form factor. In a monopole
parameterization, a cut-off
$\Lambda \approx 800\,$MeV is obtained as compared to 
$\Lambda \approx 1000\,$MeV using  $\pi\rho$ intermediate
states only.
\end{abstract}



\section{Introduction}
\vspace{-12.0cm} 
\widetext
{NT@UW-99-27, DOE/ER/40561-67, FZJ-IKP(TH)-1999-17 }
\narrowtext
\vspace{12.0cm}

The pion nucleon-nucleon vertex function is needed in many different
places in hadron physics, such as the nucleon-nucleon interaction
\cite{MHE}, pion-nucleon scattering and pion photoproduction
\cite{surya96,sato96}, deep-inelastic scattering \cite{thom83}, and as
a possible explanation of the Goldberger-Treiman discrepancy
\cite{braat72,Jones75,durso77,coon90,domin85}.  Commonly, one
represents the vertex function by a phenomenological form factor. The
cut-off parameters employed in the various calculations range from
$\Lambda^{(1)}_{\pi N N} = 300$ MeV in pion photoproduction to
$\Lambda^{(1)}_{\pi N N} = 1700$ MeV in some One-Boson Exchange models
of the nucleon-nucleon interaction, assuming a monopole
representation (i.e. $n=1$ in Eq. (\ref{ff}), see below).  
The Skyrmion model \cite{Meis86,Kai87} gives
$\Lambda^{(1)}_{\pi N N} = 860$ MeV.  A lattice gauge calculation gets
$\Lambda^{(1)}_{\pi N N} = 750 \pm 140$ MeV \cite{Liu95}.

Unfortunately, the $\pi NN$  form factor cannot be determined
experimentally. It is an off-shell quantity which is inherently
model-dependent. For a given model, 
the form factor -- besides being a parametrization of the vertex function --
summarizes those processes which are not calculated explicitly.

The simplest class of meson-theoretical models of the nucleon-nucleon
interaction includes the exchange of one meson only.
In these models, one needs "hard" form factors for the following reason.
In One-Boson Exchange potentials, the tensor
force is generated by one-pion and one-rho exchange.
 A cut-off
below 1.3 GeV would reduce the tensor force too strongly and
make a description of the quadrupole moment of the deuteron
impossible \cite{Ericson83,MHE}.

This situation changes when the exchange of two correlated bosons
between nucleons is handled explicitly.
The exchange of an interacting $\pi\rho$ pair 
generates additional tensor strength at
large momentum transfers. This implies softer cut-offs
for the genuine one-pion exchange \cite{Janssen93,Janssen94}.

Meson theory allows to undress the phenomenological form factors
at least partly by calculating those processes which contribute 
most strongly to the long range part of the vertex
 functions \cite{Janssen93,plumper}.
A physically very transparent way to include the most important
processes is given by dispersion theory.
 The imaginary part of the form factor
in the time-like region is given by the unitarity cuts.
In principle, one should consider the full three-pion continuum.
A reasonable approximation is to reduce the three-body problem
to an effective two-body problem by representing the two-pion
subsystems by effective mesons \cite{amado}. An explicit calculation
of the selfenergy of the effective meson incorporates the effects
of three-body unitarity.
In the case of the pion-nucleon form factor, one
 expects that $\pi\rho$ and $\pi\sigma$ 
intermediate states are particularly relevant.
Here,  "$\sigma$ " is understood as an abbreviation for the isoscalar
two-pion subsystem.
 In many early calculations \cite{durso77,dillig},
the effect
of the $\pi\sigma$ intermediate states was found to be 
negligible. In these calculations, 
a scalar $\sigma\pi\pi$  coupling has been used.
Nowadays, such a coupling is disfavored because it is not chirally
invariant.
In the meson-theoretical model for pion-nucleon scattering of Ref.
\cite{schutz94},
the exchange of two correlated mesons in the t-channel has been
linked to the two-pion scattering model of Ref. \cite{lohse90}.
The resulting effective  potential can be simulated
by the exchange of an effective sigma-meson in the t-channel,
if a {\it derivative} sigma two-pion coupling  is adopted.

 In the present work, we want to investigate the effect of the
 pion-sigma channel  on the pion form factor
 using a derivative coupling.

The meson-meson scattering matrix $T$ is an essential building block
of our model. Formally, the scattering matrix is obtained by 
solving the Bethe-Salpeter equation,
$ T=V+VGT, $
starting from a pseudopotential $V$.
Given the well-known difficulties in solving the
four-dimensional Bethe Salpeter equation, one rather solves
three-dimensional equations, such as
 the Blankenbecler-Sugar equation (BbS) \cite{blankenbecler}
or related ones \cite{gross,jennings}.
 The two-body propagator $G$ is chosen
 to reproduce the two-particle unitarity cuts in the physical
region. The imaginary part of $G$ is uniquely defined in this way,
but for the real part, there is  complete freedom which leads
to an infinite number of reduced equations \cite{jennings}.
The energies of the interacting particles are well-defined
for on-shell scattering only. For off-shell scattering,
there is an ambiguity.
Different choices of the energy components may affect the off-shell
behaviour of the matrix elements. As long as one is exclusively
interested in the scattering of one kind of particles, e.g. only 
pions, one can compensate the modifications of the off-shell
behaviour by readjusting the coupling constants. This gets more
difficult as soon as one aims for a consistent model of many
different reactions.  Moreover, in the calculation of the 
form factor, the scattering kernel $V$ may have singularities 
which do not agree with the physical singularities due to e.g. 
three-pion intermediate states \cite{janssen2}.

In contrast to the Blankenbecler-Sugar reduction, 
time-ordered perturbation theory (TOPT)determines the off-shell
behaviour  uniquely. Moreover, only physical
singularities corresponding to the decay into multi-meson
intermediate states can occur \cite{schweber}.
 For the present purpose, we
therefore will employ time-ordered perturbation theory.

\section{The Meson--Meson  Interaction Model}

\begin{figure}\begin{center}
\parbox{8cm}{\psfig{file=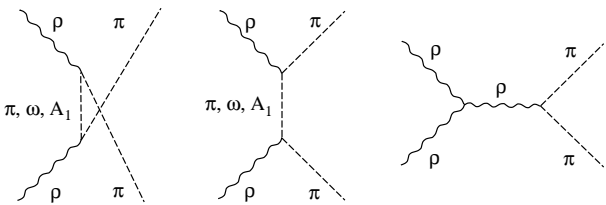,width=8cm}}
\end{center}
\caption{Diagrams describing the $\pi\rho
\rightarrow \pi\rho$ potential}
\label{rhopi}
\end{figure}

\begin{figure}\begin{center}
\parbox{8cm}{\psfig{file=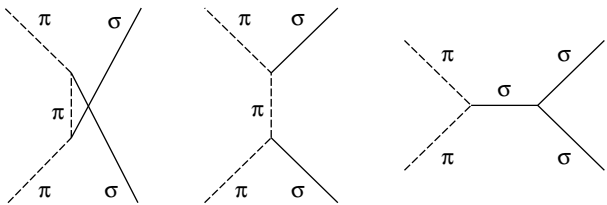,width=8cm}}
\end{center}
\caption{Diagrams describing the $\pi\sigma
\rightarrow \pi\sigma$ potential}
\label{sigpi}
\end{figure}

\begin{figure}\begin{center}
\parbox{8cm}{\psfig{file=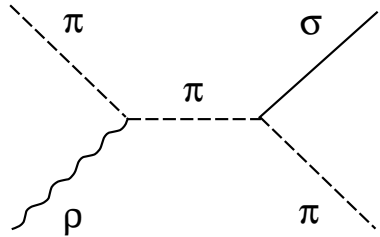,width=5cm}}
\end{center}
\caption{Diagram describing the $\pi\rho
\rightarrow \pi\sigma$ transition potential}
\label{rhosig}
\end{figure}

 The Feynman diagrams defining the pseudopotentials for
$\pi\rho$ and $\pi\sigma$ scattering are shown in Fig.  \ref{rhopi}
and Fig.  \ref{sigpi}.
We include both pole diagrams as well as t-channel and u-channel
exchanges. The transition potential is given by one-pion exchange
in the t-channel (see Fig. \ref{rhosig}).
 In Ref. \cite{Janssen93}, $\pi\rho$
scattering has been investigated neglecting the $A_1$-exchange
in the u-channel.

The $\pi\pi\rho$ and $A_1\pi\rho$ interactions are chosen according to
the Wess-Zumino Lagrangian
\cite{zumino} with $\kappa = \frac{1}{2}$.
The two-pion sigma vertex is defined by a derivative coupling \cite{schutz94}. 
For the three-sigma vertex we take a scalar coupling \cite{schutz94}. 
 The Lagrangians employed read explicitly:

\begin{eqnarray}
{\cal L}_{\pi\pi\rho} & = & -g_{\pi\pi\rho}\ (\vec{\pi} \times
\partial_{\mu} \vec{\pi}) \cdot \vec{\rho}^{\,\mu}\\ {\cal
L}_{\rho\rho\rho} & = &
\frac{1}{2}g_{\pi\pi\rho}(\partial_{\mu}\vec{\rho}_{\nu} -
\partial_{\nu}\vec{\rho}_{\mu}) \cdot (\vec{\rho}^{\,\mu} \times
\vec{\rho}^{\,\nu})\\ {\cal L}_{A_1\pi\rho} & = &
\frac{g_{\pi\pi\rho}}{m_{A_1}} \Big[(\vec{A}_{\mu} \times
\partial_{\nu}\vec{\pi}) - (\vec{A}_{\nu} \times
\partial_{\mu}\vec{\pi})\nonumber\\ & & + \frac{1}{2} \left\{\vec{\pi}
\times (\partial_{\mu}\vec{A}_{\nu} -
\partial_{\nu}\vec{A}_{\mu})\right\}\Big]
(\partial^{\mu}\vec{\rho}^{\,\nu} - \partial^{\nu}\vec{\rho}^{\,\mu})
\\ {\cal L}_{\omega\pi\rho} & = &
\frac{g_{\omega\pi\rho}}{m_{\omega}}\ \epsilon^{\mu\alpha\lambda\nu}
\partial_{\alpha}\vec{\rho}_{\mu}\ \partial_{\lambda}\vec{\pi}\
\omega_{\nu} \\
 {\cal L}_{\pi\pi\sigma} & = & \frac{f}{2m_{\pi}}\
\partial_{\mu}\vec{\pi} \cdot \partial^{\mu}\vec{\pi} {\sigma}\\
{\cal L}_{\sigma\sigma\sigma} & = & g_{\sigma\sigma\sigma} m_{\sigma}\
{\sigma}{\sigma}{\sigma}.
\end{eqnarray} 
The completely antisymmetric tensor 
has the component $\epsilon_{0123}=+1$.

In the presence of derivative couplings, the canonical momenta
conjugate to the fields $\Phi_k$,
 $$\pi_k=\frac{\delta{\cal L}}{\delta\dot{\Phi}_k},$$
receive contributions from the interaction Lagrangian. The
corresponding Hamiltonian density
\begin{equation}
{\cal H}\ =\ \sum_k \pi_k \dot{\Phi}_k\ -\ {\cal L}
\end{equation}
consists of the ususal terms plus additional contact terms:
\begin{equation}
{\cal H}\ =\  {\cal H}_0 - {\cal L}_{int} + {\cal H}_{contact}.
\end{equation}
The contact terms ensure that the diagrams of time-ordered
perturbation theory are on-shell equivalent to the corresponding
Feynman diagrams. For our model Lagrangian,
the contact terms, expanded up to the order $g^2, f^2$, and $fg$, are
given by: 
\begin{eqnarray}
{\cal H}_{contact} & = &
 +\frac{g_{\pi\pi\rho}^2}{2}(\vec{\rho}^{\,0} \times \vec{\pi})^2 +
 \frac{f^2}{2m_{\pi}^2} {\sigma}^2 \dot{\vec{\pi}}^2\nonumber\\
 & & -
 \frac{fg_{\pi\pi\rho}}{m_{\pi}} {\sigma}\dot{\vec{\pi}} \cdot
 (\vec{\rho}^{\,0} \times \vec{\pi})\nonumber\\ 
 & & +
 \frac{g_{\omega\pi\rho}^2}{2m_{\omega}^4}\left\{\epsilon_{ijk}(\partial^i
 \vec{\rho}^{\,j})(\partial^k \vec{\pi})\right\}^2\nonumber\nonumber\\
 & & +\frac{2g_{\pi\pi\rho}^2}{m_{A_{1}}^{4}}
 \left\{\partial_{\nu}\vec{\pi} \times
 (\dot{\vec{\rho}}^{\,\nu}-\partial^{\nu}\vec{\rho}^{\,0})\right\}^2\nonumber\\
 & & + \frac{g_{\pi\pi\rho}^2}{2m_{A_{1}}^{2}} \left\{(\partial_k
 \vec{\rho}^{\,0} - \dot{\vec{\rho}}_k) \times \vec{\pi}\right\}^2
\end{eqnarray}

\begin{figure}\begin{center}
{$\parbox[c][2.1cm]{1.6cm}
{\psfig{file=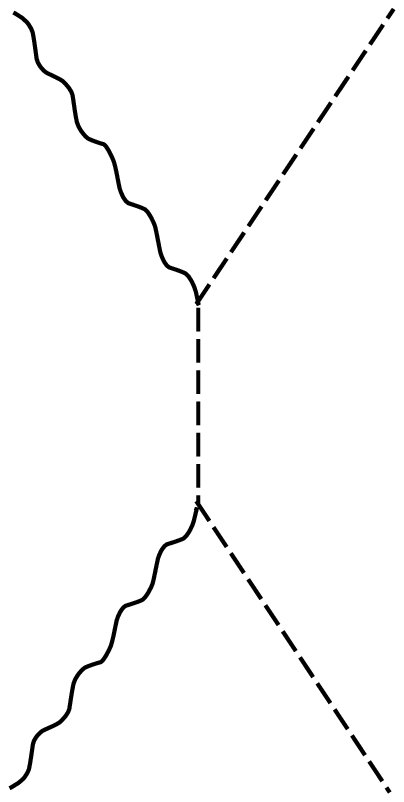,height=2cm,width=1.5cm}}
= 
\parbox[c]{1.6cm}
{\psfig{file=fig4.eps,height=2cm,width=1.5cm}}
\;+\;\parbox[c]{1.6cm}
{\psfig{file=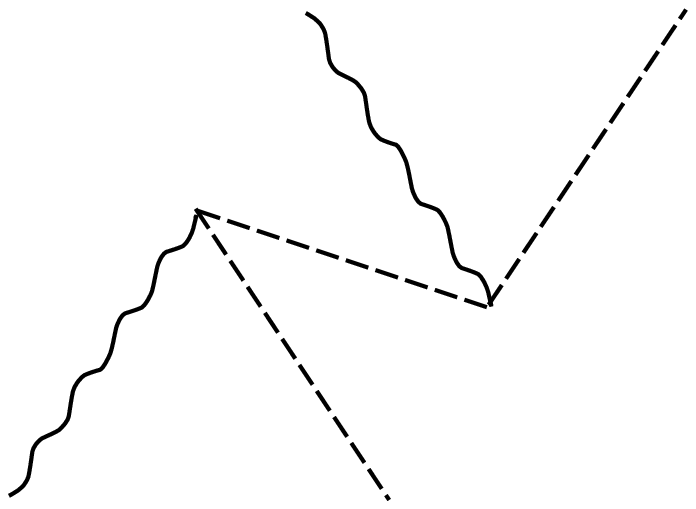,height=2cm,width=1.5cm}}
\;+\; \parbox[c]{1.6cm}
{\psfig{file=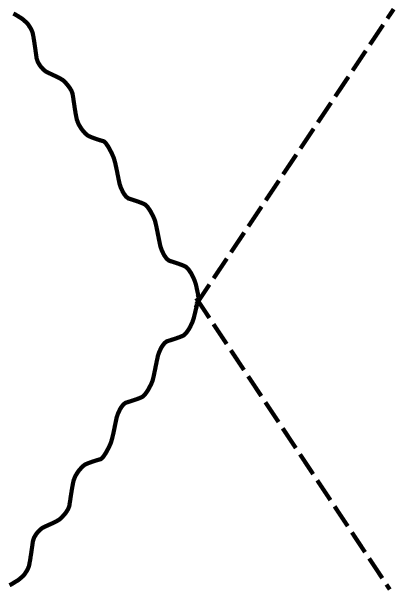,height=2cm,width=1.5cm}}
$}
\caption{Decomposition of the pion-pole Feynman diagram
into time-ordered diagrams and a contact term.}
\label{topt-bbs-aequi}
\end{center}\end{figure}

The Feynman diagrams are replaced by the corresponding time-ordered
diagrams and a contact diagram, as e.g. shown in Fig. 
\ref{topt-bbs-aequi}
for the case of pion-rho scattering via pion exchange in the
s-channel. 

Form factors are required to ensure  convergence.  We choose 
standard monopole ($n=1$) or dipole ($n=2$) parameterizations,
\begin{equation}\label{ff}
\Gamma^{(n)}(q^2) = \left(\frac{\Lambda^2-m^2}
{\Lambda^2-q^2}\right)^n 
\end{equation}
The cut-off parameters $\Lambda^{(n)}$ and the coupling constants
$\frac {g^2}{4\pi}$ are taken from other investigations. In detail,
we employ the following constants.
The coupling constant 
$\frac {g^2_{\pi\pi\rho}}{4\pi}=2.84$ can be determined from the decay 
$\rho\rightarrow \pi\pi$.
We assume $ g_{\pi\rho A_1}= g_{\pi\pi\rho}$ \cite{zumino}.
 The corresponding cut-off parameters
$\Lambda^{(1)}_{\pi\pi\rho}=1500 $ MeV and
$\Lambda^{(2)}_{\pi\rho A_1}=2600 $ MeV
 have been taken from Ref. \cite{janssen2}. 
The decay $\omega\rightarrow \pi\rho \rightarrow \pi\gamma $
gives the coupling constant
$\frac {g^2_{\pi\rho\omega}}{4\pi}=7.5 $  and 
$\Lambda^{(2)}=2200 $ MeV \cite{durso87}.

In the meson-theoretic model for pion-nucleon scattering of Refs.
\cite{schutz94,reuber}, the following constants have been determined:
$\frac {g^2_{\pi\pi\sigma}}{4\pi}=0.25$,
$\Lambda^{(1)}_{\pi\pi\sigma}=1300$ MeV,
$\frac {g^2_{\sigma\sigma\sigma}}{4\pi}=3.5$,
$\Lambda^{(1)}_{\sigma\sigma\sigma}=2000$ MeV.

\begin{figure}\begin{center}
\parbox{8cm}{\psfig{file=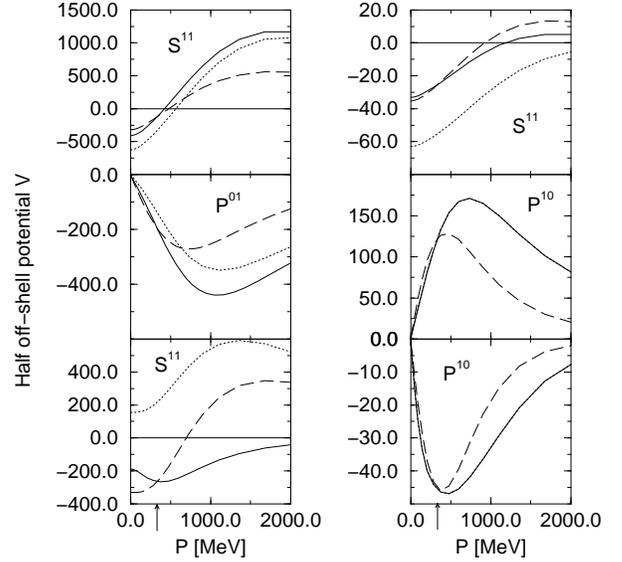,width=8cm}}
\end{center}
\caption{Different contributions to the 
half off-shell potentials for $\pi\rho$ 
scattering at $\sqrt{s}=1.2$ GeV for various partial waves as functions
of the off-shell momentum P. The arrow indicates the on-shell momentum
P corresponding to $\sqrt{s}=1.2$ GeV. The solid line represents
the scattering kernel $V$ of TOPT, while the dashed line refers to the
BbS reduction. The dotted line shows the TOPT result omitting the
contact terms. The panels show contributions of specific diagrams
of Fig.\ref{rhopi}; upper left: $A_1$ pole diagram, middle left: 
$\pi$ pole diagram, lower left: $\rho-$exchange in the t-channel,
upper right: $A_1$ u-channel exchange, middle right: $\omega$
pole diagram, lower right: $\omega$ u-channel exchange.
}\label{hos_pot}
\end{figure}

In Fig.\ref{hos_pot},
  we compare the half-off shell scattering kernels $V$ derived
in TOPT and in the BbS reduction for a 
center-of-momentum energy $\sqrt{s}=1.2$ GeV. 
When the off-shell momentum P is equal to the incoming on-shell
momentum, the potentials $V$ evaluated in the BbS reduction and
in TOPT are identical (see the arrow).
In the BbS reduction,
the zeroth component of the momentum vectors is not well-defined
for off-shell scattering. In Fig.\ref{hos_pot}
  we have chosen on-energy shell
components, following Ref. \cite{Janssen93}. For the $A_1$ exchange
(both in the s- and in the u-channel),
both formalisms give fairly similar results. For the rho-exchange
in the t-channel, the $S^{11}$ partial wave shows large differences:
while TOPT predicts an attractive half-off shell matrix element,
the BbS-reduction gets repulsive for momenta larger than 650 MeV.

The singularities in the scattering kernel $V$ due to unitarity
cuts are handled by chosing an appropriate path in the complex
momentum plane. The scattering equation, after partial wave
decomposition, reads explicitly:
\begin{equation}
T(p,p') = V(p,p') + \\
 \int dk k^2 V(p,k) G^{TOPT}(E;k) T(k,p') 
\end{equation}
with
\begin{eqnarray}
  k=|\vec{k}| e^{-i\Phi},
\end{eqnarray}
where $\Phi$ is a suitably chosen angle \cite{hether}.
  $G^{TOPT}(E; k)$ denotes the two-body propagator of time-ordered
perturbation theory.

\begin{figure}\begin{center}
\parbox{8cm}{\psfig{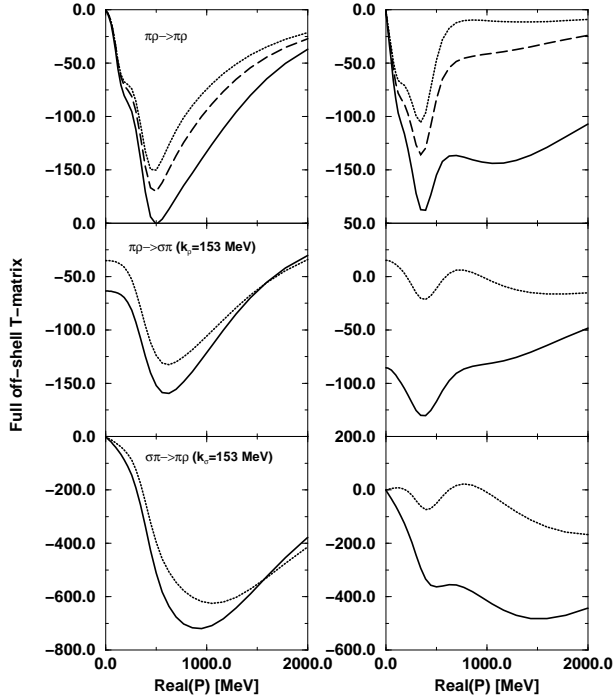}}
\end{center}
\caption{The full off-shell $T$-matrices for 
the transitions $\pi\rho\rightarrow\pi\rho$
(upper panels), $\pi\rho\rightarrow\sigma\pi$ (central panels), and
$\sigma\pi\rightarrow\pi\rho$ (lower panels) are shown for $E_{CM}=1200$MeV
and $k^{in}_{CM}=153 $ MeV(solid lines) as functions of the
real part of the complex
off-shell momentum P.
The $T$-matrix for the transition $\pi\rho\rightarrow\pi\rho$ obtained without
coupling to the
$\sigma\pi$ channel is given by the dashed line.
For comparison, also the corresponding scattering kernels $V$ are
displayed (dotted lines). The real parts of the $T$-matrices are shown
on the left hand side, the imaginary parts on the right hand side.
}\label{E1200}
\end{figure}

The resulting $T$-matrix is shown  in Fig.\ref{E1200}
  for $\sqrt{s}=0.7$ GeV and k=153 MeV for
the total angular momentum J=0. The potential $V$ for pion-rho
scattering is  attractive (upper panel).
 Iterating the $\pi\rho$ diagrams by
themselves, the attraction is enhanced. The inclusion of the
$\pi\sigma$ channel enhances the attraction even more.
This effect is due  to the off-shell transition potential
(middle and lower panel) which is  larger than the diagonal $\pi\rho$
potential. The magnitude of the transition potential
can be traced back to the interaction Lagrangian
(see Fig.\ref{hos_pot}).
 The derivative
coupling favours large momentum transfers.  
The enhancement of the $\pi \rho - \pi \rho $ scattering matrix
$T$ due to these coupled channel effects will shift the maximum
of the spectral function to lower energies and thus lead to a
softer form factor. 

\section{The Pion--Nucleon   Form Factor}

\begin{figure}\begin{center}
\parbox{8cm}{\psfig{file=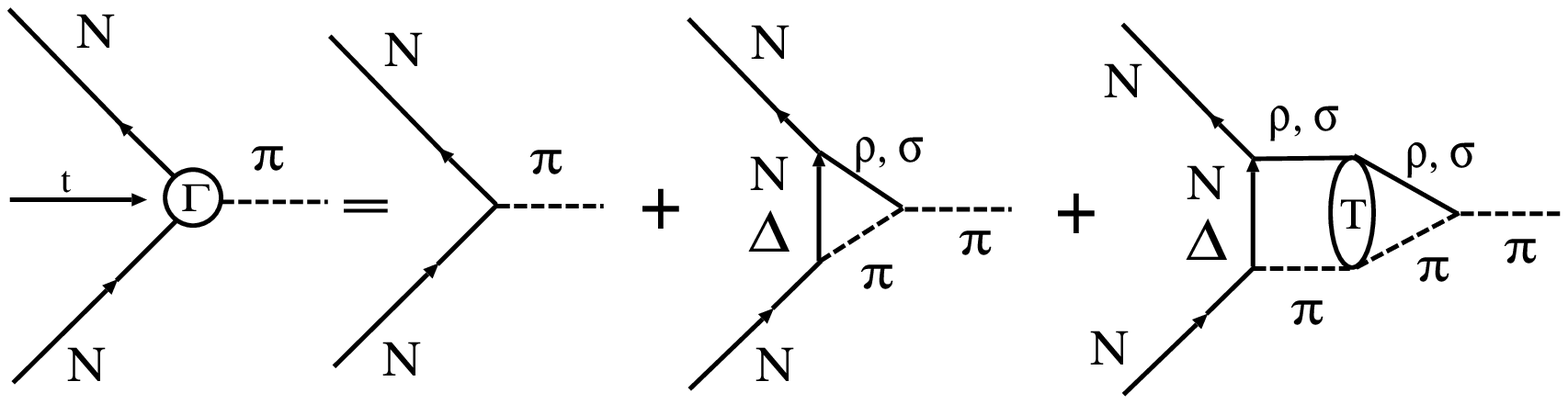,width=8cm}}
\end{center}
\caption{The $\pi NN$ vertex function}\label{Gam_piNN}
\end{figure}

The present model for the $\pi NN$ vertex function F  is shown in
Fig.  \ref{Gam_piNN}.

 Before coupling to the nucleon, the pion can disintegrate into
three-pion states which are summarized by both pion-sigma and 
pion-rho pairs. We first evaluate the vertex function in the
$N \bar N $ channel. In this channel, the $\pi\rho$ and
$\pi\sigma $ interactions can be summed.  After a decomposition
into partial waves, one gets:

\begin{eqnarray}
  F_{N\bar{N}\rightarrow\pi}  =  F^{\,0}_{N\bar{N}\rightarrow\pi}
 + \sum_{n=\rho,\,\sigma}\int dk\,k^2
\nonumber \\ 
\times f_{\pi\leftarrow\pi n}(t,k) G_{\pi n}(t,k) V_{\pi
    n \leftarrow N\bar{N}}(t;k,p_0)\;.
\end{eqnarray}
Here, $\rm p_0$ is the subthreshold
{\it{on--shell}} momentum of the $\rm
N\bar{N}\,$--System \cite{braat72}.
The bare vertex is called
  $F^{\,0}_{N\bar{N}\rightarrow\pi}$.
 We have to include the self energies $\Sigma_\rho$ and
$\Sigma_\sigma$ of both the $\rho$ and the $\sigma$
into the two-particle propagators $\rm G_{\pi\rho}$ and
$\rm G_{\pi\sigma}$ 
 because the vertex function is needed in the time-like region.
The annihilation potential
 $\rm V_{\pi n \leftarrow
  N\bar{N}}$ has been worked out in Ref. \cite{janssen4}.
The form factor needed for the $N\bar{N}\rightarrow\pi\rho(\sigma)$
transition has not been determined selfconsistently, but taken
from Ref. \cite{janssen4}.

The dressed 
meson--meson $\rightarrow$ pion vertex function
$ f_{\pi\leftarrow\pi n}$ is given by

\begin{eqnarray}
  f_{\pi\leftarrow\pi n}(t,k) = f^{\,0}_{\pi\leftarrow\pi n}(t,k)
   + \sum_{m=\rho,\,\sigma} \int dk' {k'}^2
\nonumber
\\
 \times f^{\,0}_{\pi\leftarrow\pi
    m}(t,k') G_{\pi m}(t,k') T_{\pi m\leftarrow \pi n}(t;k',k)\;.
\end{eqnarray}
 The bare vertex function is called $ f^{\,0}$. 
The vertex function $f$ requires the off-shell elements of the $T$-matrix
 for meson-meson scattering
 $ T_{\pi m\rightarrow \pi n}$ 
discussed in the previous chapter.
Only the partial wave with total angular momentum
 $\rm J^\pi=0^-$
of the
 $ N\bar{N}\rightarrow\pi$ vertex function is needed.

The form factor
 $\rm \Gamma(t)$ is defined as
\begin{eqnarray}
  \Gamma(t) & = &
  \frac{F_{N\bar{N}\rightarrow\pi}}{F^{\,0}_{N\bar{N}\rightarrow\pi}}\;.
\end{eqnarray}

Now we rely on dispersion theory to obtain the form factor
$\rm \Gamma(t)$ for space-like momentum transfers t
\cite{braat72,durso77}.
We employ a subtracted dispersion relation
\begin{eqnarray}
  \Gamma(t) & = & 1+\frac{t-m_{\pi}^2}{\pi} \int\limits_{9m_{\pi}^2}^{\infty}
  \frac{Im\Gamma(t')dt'}{(t'-t)(t'-m_{\pi}^2)}\;.\label{subintegral}
\end{eqnarray}
The subtraction constant ensures that the form factor is normalized
to unity for $t=m_{\pi}^2$.
The integration is cut off at
 $ t=4\,m_N^{\,2}$. Larger values of $t$ would require to 
incorporate diagrams with cuts at larger energies. Such processes
are suppressed because of the improved rate of convergence
of the subtracted dispersion relation.

\begin{figure}\begin{center}
\parbox{8cm}{\psfig{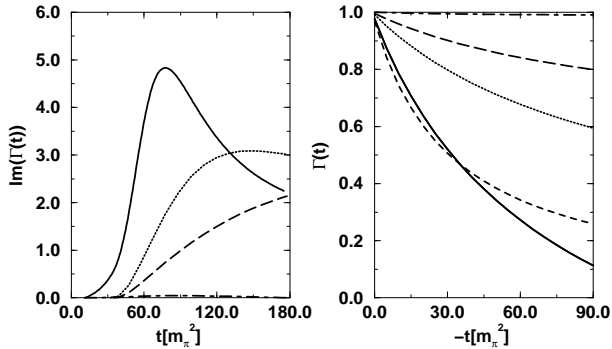}}
\end{center}
\caption{Real (right panel) and imaginary (left panel) parts
of the $\pi NN$ form factor as functions of the momentum transfer t.
Solid line: coupled $\pi\rho$ and $\pi\sigma$ channels;
dotted line: only the $\pi\rho$ channel is considered;
dashed line: rescattering in the $\pi\rho$ channel is omitted;
dashed-dotted line: only the $\pi\sigma$ channel is included.
The short-dashed line in the right panel refers to a monopole
form factor with $\Lambda^{(1)}=800$ MeV.
}\label{spek_Gam}
\end{figure}

The imaginary part of the form factor is shown in the left part of
 Fig.\ref{spek_Gam}.

We confirm earlier findings \cite{durso77,dillig}
 that the pion-sigma states by themselves,
even if iterated, do not generate  an appreciable contribution
to the spectral function. The $\pi \rho$ intermediate states clearly
dominate. Including rescattering processes in a model with 
only $\pi\rho$ states, one finds a large shift of the spectral
function to smaller energies, which emphasizes the importance of the
correlations. The new aspect of our work is the large shift induced
by the coupling between $\pi \rho $ and $ \pi \sigma $ states.

In Ref. \cite{Hol} Holinde and Thomas introduced an effective $\pi'$
exchange contribution to the One-Boson Exchange $NN$ interaction in a
phenomenological way in order to shift part of the tensor force from
the $\pi$ into the $\pi'$ exchange. This allowed them to use a rather
soft cut-off $\Lambda^{(1)}_{\pi NN}=800 MeV$ to describe the $NN$
phase shifts.  The maximum of the spectral function shown in Fig.
\ref{spek_Gam} is located at 1.2 GeV ($t \approx 75 m_{\pi}^2$) which
coincides with the mass of the $\pi'$. Thus the $\pi'$ used in Ref.
\cite{Hol} can be interpreted in terms of a correlated coupled
$\pi\rho$ and $\pi\sigma$ exchange.  Note, that the form factor
derived here must not be used in Meson-Exchange models of the
nucleon-nucleon interaction, such as discussed in Ref. \cite{MHE}, but
only in models which include the exchange of {\it correlated}
$\pi\rho$ and $\pi\sigma$ pairs.

%

The form factors obtained via the dipersion relation are shown
in the right part of Fig.\ref{spek_Gam}.
The numerical results can be parameterized by a monopole form
factor.
The inclusion of an uncorrelated $\pi \rho$ exchange leads to a   
relatively hard form factor of 
 $\Lambda^{(1)}_{\pi N N} = 2100$ MeV.
In the present model, using $\pi\rho$ intermediate states only, the
cut-off momentum is reduced to
 $\Lambda^{(1)}_{\pi N N} = 1500$ MeV.
If one treats the singularities of the scattering kernel $V$ 
by approximating the propagator of the virtual pion by a static
one (see the u-channel exchange in Fig.  \ref{rhopi}), the 
resulting $\pi\rho$ interaction becomes more attractive and
produces a much softer form factor corresponding to
$\Lambda^{(1)}_{\pi N N} = 1000$ MeV \cite{Janssen93}, employing
only $\pi\rho$ intermediate states.
The present model, including both $\pi\rho$ and $\pi\sigma$
intermediate states, 
leads to $\Lambda^{(1)}_{\pi N N} = 800$ MeV.

\section{Summary}
Microscopic models of the $\pi NN$ vertex function are required
in order to understand why the phenomenological form factors employed  
in models of the two-nucleon interaction are harder than those
obtained from other sources. In Ref. \cite{Janssen93}, a 
meson-theoretic model for the $\pi NN$ vertex has been developped.
The inclusion of correlated $\pi\rho$ states gave a form factor
corresponding to $\Lambda^{(1)}_{\pi N N} = 1000$ MeV.
This is still harder than the phenomenological form factors 
required in the description of many other physical processes.
 Within
the framework of Ref. \cite{Janssen93}, a further
 reduction of the cut-off $\Lambda^{(1)}_{\pi N N}$
is impossible. Correlated $\pi\sigma$ states were not considered
in Ref. \cite{Janssen93} because of the results obtained in
Refs. \cite{durso77,dillig}. In the present work we have shown
that the findings of Refs. \cite{durso77,dillig} have to be 
revised. Meson-theoretic analyses of $\pi N$ scattering
strongly suggest a derivative $\sigma\pi\pi$ coupling.
This is shown to enhance the off-shell coupling between
$\pi\rho$ and $\pi\sigma$ intermediate states in the dispersion
model for the $\pi NN$ form factor. A softening of the
$\pi NN$ form factor is obtained which largely removes the
remaining discrepancies between the phenomenological
form factors.

{\bf Acknowledgments}

C.H. acknowledges the financial support through a Feodor-Lynen
Fellowship of the Alexander-von-Humboldt Foundation. This work was
supported in part by th U.S. Department of Energy under Grant No.
DE-FG03-97ER41014.



%
%

%
%


\begin{references}

\bibitem{MHE} R.~Machleidt, K.~Holinde and C.~Elster, 
Phys. Rep. {\bf 149}, 1 (1987).

\bibitem{surya96}Y.~Surya and F.~Gross,
Phys. Rev. C {\bf 53}, 2422 (1996).

\bibitem{sato96}T.~Sato and T.-S.H.~Lee,
Phys. Rev. C {\bf 54}, 2660 (1996).


\bibitem{thom83}A.W.~Thomas, Phys. Lett. B {\bf126}, 97 (1983).

\bibitem{braat72}H.J.~Braathen,
Nucl. Phys. {\bf B44}, 93 (1972).

\bibitem{Jones75}H.F.~Jones and M.D.~Scadron,
Phys. Rev. D {\bf 11}, 174 (1975).

\bibitem{durso77} J. W.~Durso, A. D.~Jackson and B. J.~Verwest, 
Nucl. Phys. {\bf A282} (1977) 404. 

\bibitem{coon90}S.A.~Coon and M.D.~Scadron,
Phys. Rev. C {\bf 42}, 2256 (1990).

\bibitem{domin85}C.A.~Dominguez,
Nuovo Cimento {\bf 8}, 1 (1985).




\bibitem{Meis86}U.G.~Mei\ss{}ner et al.
Phys. Rev. Lett. {\bf 57}, 1676 (1986). 

\bibitem{Kai87}N.~Kaiser,U.G.~Mei\ss{}ner, and W.~Weise,
 Phys. Lett. {\bf B198}, 319 (1987).

\bibitem{Liu95} K. F. ~Liu, S. J. ~Dong, T. ~Draper and W. ~Wilcox, 
Phys. Rev. Lett. {\bf 74}, 2172 (1995). 

\bibitem{Ericson83}T.E.O.~Ericson and M.~Rosa-Clot,
Nucl. Phys. {\bf A405}, 497 (1983).

\bibitem{Janssen93} G. ~Janssen, 
J. W. ~Durso, K. ~Holinde, B.C. ~Pearce and J. ~Speth,
Phys. Rev. Lett. {\bf 71}, 1978 (1993).

\bibitem{Janssen94} G.~Janssen, 
 K.~Holinde and J.~Speth,
Phys. Rev. Lett. {\bf 73}, 1332 (1994).


\bibitem{plumper} D.~Pl\"umper, J.~Flender, and M.~Gari,
Phys. Rev. C {\bf 49}, 2370 (1994).

\bibitem{amado}R.~Aaron, R.D.~Amado and J.E.~Young,
Phys. Rev. {\bf 174}, 2022 (1968).

\bibitem{dillig} M. ~Dillig and M. ~Brack,
J. Phys. G {\bf 5}, 233 (1979).

\bibitem{schutz94} C.~Sch\"utz, J.W.~Durso, K.~Holinde, and J.~Speth,
Phys. Rev. C {\bf 49}, 2671 (1994).

\bibitem{lohse90} 
D.~Lohse, J.W.~Durso, K.~Holinde, and J.~Speth,
Nucl. Phys. {\bf A516}, 513 (1990).

\bibitem{blankenbecler} R.~Blankenbecler and R.~Sugar,
Phys. Rev. {\bf 142}, 1051 (1966).

\bibitem{gross} F.~Gross,
Phys. Rev. C {\bf 26}, 2203 (1982).

\bibitem{jennings} E.D.~Cooper and B.K.~Jennings,
Nucl. Phys. {\bf A500}, 553 (1989).

\bibitem{janssen2} G.~Janssen, K.~Holinde and J.~Speth,
Phys. Rev. C {\bf 49}, 2763 (1994).

\bibitem{schweber} S.S.~Schweber, {\it An Introduction to
relativistic Quantum Field Theory},
(Harper and Row, 1962).

\bibitem{zumino} J.~Wess and B.~Zumino, 
Phys. Rev. {\bf 163}, 1727 (1967).

\bibitem{durso87} J.W.~Durso,
Phys. Lett. {\bf B184}, 348 (1987); 
G.~Janssen, J\"ul-report 2734, (1993).

\bibitem{reuber} A.~Reuber, K.~Holinde, H.C.~Kim, and J.~Speth,
Nucl. Phys. {\bf A608}, 243 (1996).

\bibitem{hether}J.H.~Hetherington and L.W.~Schick, 
Phys. Rev. B {\bf 137}, 935 (1965).

\bibitem{janssen4} G. ~Janssen, K. ~Holinde and J. ~Speth,
Phys. Rev. C {\bf 54}, 2218 (1996).

\bibitem{Hol} K. ~Holinde and A. W. ~Thomas, 
Phys. Rev. C {\bf 42} (1990) R1195.

\end{references}
\end{document}